\documentstyle[preprint,revtex]{aps}
\begin{document}
\draft
\preprint{Princeton-Trieste-Rutgers-9305}
\preprint{April 1993}
\begin{title}
Gutzwiller-Jastrow Wavefunctions for the $1/r$ Hubbard Model
\end{title}
\author{D. F. Wang$^{\dagger}$}
\begin{instit}
Joseph Henry Laboratories of Physics, Princeton University, \\
Princeton, New Jersey 08544
\end{instit}
\author{Q. F. Zhong}
\begin{instit}
International School for Advanced Study\\
Via Beirut 4, 34014 Trieste, Italy
\end{instit}
\author{P. Coleman}
\begin{instit}
Serin Physics Laboratory, Rutgers University\\
P. O. Box 849, Piscataway, NJ 08854
\end{instit}
\begin{abstract}
In this work, we study the wavefunctions of the one dimensional $1/r$ Hubbard
model in the strong interaction limit $U =\infty$. A set of Gutzwiller-Jastorw
wavefunctions are shown to be eigen-functions of the Hamiltonian.
The entire excitation spectrum and
the thermodynamics are also studied in terms of more generalized
Jastrow wavefunctions.
For the wavefunctions and integrability conditions
at finite on-site energy, further investigations are needed.
\end{abstract}

%%%%%%%%%%%%%%%%%%%%%%%%%%%%%%%%%%%%%%%%%%%%%%%%%%%%%%%%%%%%%%%%
\narrowtext
%\section{INTRODUCTION}
There have been considerable interests in the study of low
dimensional integrable
models\cite{arovas,haldane,shastry,poly,min,kuramoto,kawakami,sutherland,Rucken91,anderson89,ha92,wang,coleman,sutherland3,sutherland2,Frahm}.
One of them is the Hubbard model with $1/r$
hopping and $U$ on-site energy, introduced by
Gebhard and Ruckenstein\cite{Rucken91}. In this work we report some new
results in our recent study of the system.

The long range Hubbard model is exactly solvable in one dimension
for arbitrary on-site energy\cite{Rucken91}.
Based on small size numerical work and several limiting cases, Gebhard and
Ruckenstein have introduced an effective
Hamiltonian, which was conjectured to be equivalent in describing
the Hubbard model for arbitrary on-site energy.
With the effective Hamiltonian, the system has been found to
exhibit a metal-insulator transition when the bandwidth $t$ is
equal to the on-site energy $U$ in the half filling case, and
the Gutzwiller state is an eigen-state of the effective Hamiltonian in
the large $U$ limit.
The complete wavefunctions and the integrability conditions have remained
unsolved for a long time.

In this primary report, we only concentrate on the study of
the wavefunctions and the spectrum
in the strong interaction limit $U=\infty$.
In this extreme limit, one unfortunate artifact is that the spins
are completely free and decoupled trivially from the charge degrees
of freedom.
We introduce a set of generalized Gutzwiller-Jastrow wavefunctions, and
we show that they are exact eigen-functions of the
Hamiltonian. The lowest energy state in this set is also a ground
state of the system in the whole Hilbert space. Considering more
generalized Jastrow wavefunctions, we are able to write the full
spectrum.  The separation of spin and charge in the full
excitation spectrum shows that the system is a special
example of Luttinger liquids in the sense of Haldane.
%%%%%%%%%%%%%%%%%%%%%%%%%%%%%%%%%%%%%%%%%%%%%%%%%%%%%%%%%%%%%%%%%%%%%%%

The Hamiltonian for the one-dimensional Hubbard model is given by
\begin{equation}
H=\sum_{i\ne j;\sigma=\uparrow,\downarrow} t_{ij} c_{i\sigma}^{\dagger}
c_{j\sigma} + U \sum_{i}
n_{i\uparrow} n_{i\downarrow},
\label{eq:hamil}
\end{equation}
where $c_{i\sigma}^{\dagger}$ and $c_{i\sigma}$ are creation and
annihilation operators at site $i$ with spin component $\sigma$.
We take $t_{ij}=it (-1)^{(i-j)}
/d(i-j)$ where $d(n)={L\over\pi}\sin(n\pi/L)$
is the chord distance\cite{Rucken91}.
Here we assume periodic boundary condition for the wavefunctions
for odd $L$, or anti-periodic boundary condition for even $L$.

In the strong interaction limit $U=\infty$, each site can
be occupied by at most one electron. In this limit, the Hamiltonian
can be written in terms of the Hubbard operator as follows:
\begin{equation}
H= \sum_{i\ne j; \sigma=\uparrow,\downarrow}
t_{ij} X_i^{\sigma 0} X_j^{0\sigma}.
\end{equation}
In the following, where-ever in case of the strong interaction, we always
implicitly assume no double occupancy. Let us denote the number of holes
by $Q$, the number of down-spins by $M$.
Following notations used in previous literatures,
the state vectors can be represented by creating spin and
charge excitations from the
fully polarized up-spin state $|P>$\cite{anderson89},
\begin{equation}
|\Phi> = \sum_{\alpha, j} \Phi (\{x_{\alpha}\},\{y_j\})
\prod_{\alpha} b_{\alpha}^{\dagger}
\prod_{j} h_j^{\dagger} |P>,
\end{equation}
where $b_{\alpha}^{\dagger} =c_{\alpha\downarrow}^{\dagger}
c_{\alpha\uparrow}$ is the operator to creat a down-spin at site
$\alpha$, and
$h_j^{\dagger} = c_{j\uparrow}$ creats a hole at site $j$.
The amplitude $\Phi (\{x_\alpha\},\{y_j\})$ is symmetric in
the down-spin positions, and antisymmetric in hole
positions.

To describe uniform motion and magnetization, we consider the
following generalized Gutzwiller-Jastrow wavefunctions,
\begin{eqnarray}
&&\Phi (x, y; J_s, J_h) = \exp {2\pi i\over L} (J_s\sum_{\alpha}
x_{\alpha} +J_h \sum_i y_i) \times \Phi_0\nonumber\\
&&\Phi_0 =\prod_{\alpha<\beta} d^2(x_\alpha -x_\beta)
\cdot \prod_{\alpha i} d(x_\alpha -y_i) \cdot \prod_{i<j}
d(y_i-y_j),
\end{eqnarray}
where the quantum numbers $J_s$ and $J_h$ govern the momenta of
the down-spins and holes, respectively. They can be integers or half
integers so that we have appropriate periodicities (or
anti-periodicities)
for the wavefunctions under the translations
$x_\alpha \rightarrow x_\alpha +L$, or $y_i \rightarrow y_i +L$ for
odd $L$ (or even $L$).
For the wavefunctions to be exact eigen-functions of the Hamiltonian,
the momenta of down-spins and holes must take values from some
restricted regions, which will be specified below.

These Gutzwiller-Jastrow wavefunctions have been studied extensively
in the recent study of the integrable models of long range
interaction\cite{kuramoto,kawakami,anderson89,ha92,wang}.
They were found to be the exact eigen-functions of those
systems. In the following, we demonstrate that they are also the
exact solutions of the Hubbard model.
To show that they are eigen-functions of the Hamiltonian, we
have to treat the hopping operator very carefully.

The hopping operator
can be broken up into two parts, the up-spin hopping operator and
the down-spin hopping operator.
For the up-spin hopping operator $T_{\uparrow} = \sum_{i \ne j}
t_{ij} c_{i\uparrow}^\dagger c_{j\uparrow}$, its effect
is just the hopping of holes alone when it operates on the wavefunctions.
But the
down-spin hopping operator, $T_{\downarrow}=\sum_{i \ne j}
t_{ij} c_{i\downarrow}^{\dagger} c_{j\downarrow}$,
will involve the hopping of down-spins and
holes simultaneously, which needs to be treated using the spin-rotated
version of the Gutzwiller-Jastrow wavefunctions, developed in
the recent work on the supersymmetric t-J model\cite{wang}.

For the up-spin hopping, we have
\begin{equation}
{T_{\uparrow} \Phi (x,y;J_s,J_h) \over \Phi (x,y;J_s,J_h)}
= -it \sum_i \sum_{n=1}^{L-1} {(-1)^n \over d(n)}
z^{nJ_h} \prod_{j (\ne i)} F_{ij}(n) \prod_{\alpha} F_{i\alpha}(n)
\end{equation}
where $z=\exp{2\pi i\over L}$, $F_{ij}(n) = \cos(\pi n/L)
+\sin(\pi n/L) \cot\theta_{ij}$, $F_{i\alpha}(n) =
\cos(\pi n/L) +\sin (\pi n/L) \cot\theta_{i\alpha}$, with
$\theta_{ij}=\pi (y_i - y_j) /L, \theta_{i\alpha} =\pi (y_i - x_\alpha)/L$.
The sum can be calculated in a standard way,
by expanding the products and classifying the terms by the number of
particles involved. The terms with more than two particles vanish,
yielding the following result,
\begin{equation}
{T_{\uparrow} \Phi (x,y;J_s,J_h) \over \Phi (x,y;J_s,J_h)}
=- (2\pi t/L) J_h Q + (2\pi t/L) i \sum_{\alpha, i} \cot \theta_{
i \alpha}
\end{equation}
where the hole momenta satisfy the condition
$|J_h| \le L/2 -(M+Q)/2$.

To deal with the down-spin hopping operators, using the spin
rotated version of the Gutzwiller-Jastrow wavefunctions, we find
\begin{equation}
{T_{\downarrow} \Phi (x,y;J_s,J_h) \over \Phi (x,y;J_s,J_h)}
= -it \sum_i \sum_{n=1}^{L-1} {(-1)^n \over d(n)}
z^{n{\tilde J_h}} \prod_{j(\ne i)} F_{ij}(n) \prod_{\mu} F_{i\mu}(n)
\end{equation}
where $F_{i\mu}(n) =
\cos(\pi n/L) +\sin (\pi n/L) \cot\theta_{i\mu}$, $\tilde J_h =
J_h- J_s +L/2$, $\theta_{i\mu}=\pi (y_i -u_{\mu})/L$,
for $\mu =1, 2, \cdots, L-M-Q$. Here
$u_1, u_2, \cdots, u_{L-M-Q}$ are the positions of the up-spins
on the lattice. Finally, we obtain
\begin{equation}
{T_{\downarrow} \Phi (x,y; J_h, J_s) \over \Phi (x,y;J_h,J_s)}
=-(2\pi t/L) {\tilde J_h} Q + (2\pi t/L) i \sum_{\mu, i} \cot \theta_{
i \mu}.
\end{equation}
Here $|{\tilde J_h} | \le L/2 -({\tilde M}
+Q)/2$, ${\tilde M} = L-Q-M$.

After adding the spin-up and spin-down hopping operator effects together,
the two particle terms vanish due to the fact that the down-spin
electrons, the up-spin electrons and the holes span the entire lattice.
As a result, we see that the Gutzwiller-Jastrow wavefunctions
are eigenstates of the Hamiltonian, with eigen-energies given by
\begin{equation}
E(J_s,J_h)=-(2\pi  t/L) [2 J_h -J_s +L/2] Q.
\end{equation}

If we take $t$ to be positive, the lowest energy in this set
of Jastrow wavefunctions is obtained
when $J_h$ and $\tilde J_h$ reach their allowed maximum possible values,
which occurs at
$J_h = L/2 - (M+Q)/2$ and $J_s=L-M -Q/2$. Thus the dependence of ground
state energy on the number of holes $Q$ is given by:
\begin{equation}
E_0 =- (2\pi t  /L) [L/2 -Q/2] Q.
\end{equation}
This energy is also the
absolute ground state energy of the system in the whole Hilbert space,
as indicated by our numerical results of 6 site and 8 site lattices.
For this ground state wavefunction, the exponents of the
long distance behaviors of
various correlators can be obtained from the recent work by Kawakami.
In general, the ground state is not unique, and this Jastrow
wavefunction ground state is only one of them. The ground state
degeneracy is given by $(M +{\tilde M})!/(M! {\tilde M}!)$.

As seen above, the energy consists of decoupled down-spin contribution
and the hole contribution.
In principle, we may find more generalized Jastrow wavefunctions
as eigen-functions of the Hamiltonian.
As will be seen below, in our case, we may even
write the entire spectrum of the system
in the full Hilbert space in terms of a set of Jastrow functions.

To study other excitations, let us consider more generalized
Jastrow wavefunctions in the following form
\begin{equation}
\phi = \phi_s (X, Y) \phi_h (Y) \times \Phi_0.
\end{equation}
Here the functions $\phi_s$ and $ \phi_h$ are polynomials of
$X=\{\exp(2\pi i x_{\alpha}/ L)\}, Y=\{\exp (2\pi i y_i/ L)\}$.
They are totally symmetric
in their arguments, respectively.
The degrees of these polynomials must satisfy some specific conditions,
so that many particle terms vanish when applying the hopping
operators. The eigen-energy equation thus reduces to
\begin{equation}
-(2\pi t /L) [\sum_{i=1}^Q \partial_i (\phi_s \phi_h)
+ \sum_{i=1}^Q \phi_s (\partial_i+L/2) \phi_h] = E \phi_s \phi_h,
\end{equation}
where $\partial_i = Y_i\partial/\partial Y_i$. This eigen-value equation
can be solved exactly, yielding the spectrum given by
\begin{equation}
E =-(2\pi t /L) [  \sum_{i=1}^Q n_i + \sum_{\mu = 1}^{Q}
m_{\mu}].
\end{equation}
Here, the integers (or half integers ) satisfy the conditions
$|n_i| \le L/2 -({\tilde M} +Q)/2, |m_{\mu} |
\le L/2 -(M+Q)/2$, where $n_i \le n_{i+1}$ and $m_{\mu} \le m_{\mu +1}$.
This result shows that the spectrum is invariant
when changing the sign of $t$.

We may write the spectrum in terms of a set of conjugate
quantum numbers defined by $K_i = n_i +m_i +(L-Q)/2 +i$.
The spectrum is given by
\begin{equation}
E = - (2\pi t /L) \sum_{i=1}^Q K_i
+ (\pi t Q /L) (L+1),
\end{equation}
where $ K_i$ takes values from $(1, 2, \cdots, L)$. They may
be regarded as the momenta of quasi-particles ``holons''. Our numerical
results of 6 site and 8 site lattices indicate that the above
spectrum spans the full spectrum in the entire Hilbert space.

In the spectrum formula, each energy level is determined by a charge
configuration, such as $(101010)$ for $L=6$, $Q=3$, where
the 1's represent the charge momenta. We may regard the empty values
as the momenta for the quasi-particles of spin degrees.
Our numerical result shows that the degeneracy of each energy level
is given by the number of the ways to
distribute $L-Q-M$ spins $ s=+{1 \over 2}$ and $M$ spins
$s=-{1\over 2}$
among the empty values.
Thus we see that the
spin degree is decoupled from the charge degree of freedom for the
entire excitation spectrum.

For fixed number of electrons $N_e = L-Q$ on the lattice,
the free energy consists of two parts,
$F=F_1 - T N_e \ln 2$, where the second term comes
from the decoupled spin degrees of freedom, and $F_1$ is the contribution
from the charge degree of freedom, which is that of
$Q$ spinless fermions with the relativistic spectrum.
In the grand canonical case\cite{yang71,taka71,babu83},
denoting the chemical
potential of the electrons by $\mu$, we find that the
free energy per lattice site is given by
\begin{equation}
F(T,\mu)/L = -\mu -{T \over 2 \pi}
\int_{-\pi}^{\pi} dq \ln [2+ e^{\beta (q t-\mu)}].
\end{equation}
Here the number of electrons as a function of the chemical potential is
given by $N_e = - {\partial F(T,\mu) \over \partial \mu}$. We have
also checked that the free energy correctly reproduces the first
three terms in the high temperature perturbation expansion.

We have seen that the spin degrees of freedom decouple from
the charge degrees of freedom for this system in the full excitations
in the strong interaction limit. One similar system has been studied
by one of the authors. He has introduced another
integrable model, the multi-component Hubbard model of
$1/r$ hopping and $U$ on-site energy, which is the direct
generalization of the model to the $SU(N)$ case, where a set of
generalized $SU(N)$ Gutzwiller-Jastrow wavefunctions has been
shown to be exact eigen-functions of the Hamiltonian in the strong
interaction limit\cite{coleman}.

In summary, we have studied the wavefunctions of
the one dimensional $1/r$ Hubbard model in the limit
of strong interaction $U=\infty$. We have shown that
a set of Gutzwiller-Jastrow wavefunctions
are the eigenstates of the Hamiltonian, and the lowest energy in this set is
the ground state energy in the whole Hilbert space.
The full spectrum can be written in terms of more generalized Jastrow
wavefunctions.
The spin-charge separation occurs in the
full excitation spectrum,
and the system may be regarded as a special
Luttinger liquid.

Finally, we would like to stress
that the model is completely integrable for arbitrary
on-site energy $U$\cite{Rucken91}.
The most interesting thing is to
investigate the wavefunctions and the integrability conditions for
this $1/r$ Hubbard model at finite on-site energy $U$.
It would be also of great interest to study the ground state
properties, such as the spin and charge
susceptibilities, and various ground state correlation exponents
dependence on the on-site energy $U$.
For these finite on-site energy studies, further investigations
are needed.

This work was supported in part by
the National Science Foundation
under Grant NSF-DMR-89-13692.

${\dagger}$ Address after September 1993:
Institut de physique th\'eorique, \'Ecole Polytechnique F\'ed\'erale
de Lausanne, PHB-Ecublens, CH-1015 Lausanne-Switzerland.
%%%%%%%%%%%%%%%%%%%%%%%%%%%%%%%%%%%%%%%%%%%%%%%%%%%%%%%%%%%%%%%%%%%%%%%%%%%

%\figure{Low-lying energy levels of the 10 site 2 hole system from exact
%diagonalization.  The numbers associated with each state list the
%spin degeneracies starting with spin 0 on the left.
%For example, the number ``331'' indicates that we have 3 states with
%$S=0$,  3 states with $S=1$ and 1 state with $S=2$.
%\label{figone}
%}
\end{document}